\documentclass[letterpaper,twocolumn,10pt]{article}
\usepackage{usenix}
\usepackage{graphicx} 
\usepackage{amsfonts}
\usepackage{subcaption}
\usepackage{amssymb}
\usepackage{algorithm}
\usepackage{algpseudocode}
\usepackage{pifont}   
\newcommand{\system}{\textsc{DNS-CPM}}
\newcommand{\systemFull}{DNS Cache Poisoning Mitigation}
\usepackage{amsmath}
\usepackage{xcolor}
\usepackage{graphicx}
\usepackage{pgfplots}
\usepackage{booktabs}
\usepackage{multirow}
\usepackage{enumerate}    
\usepackage{url} 
\newcommand{\xmark}{\ding{55}}    

\newcommand{\ignore}[1]{}

\newcommand{\Yehuda}[1]{\textcolor{black}{#1}}
\newcommand{\Harel}[1]{\textcolor{black}{#1}}

\begin{document}
%
\title{\system: DNS Cache Poisoning Mitigation}

 \author{
 Yehuda Afek \\
    Tel Aviv University \\
    \texttt{afek@tauex.tau.ac.il}
    \and
    Harel Berger \\
    Georgetown University \\
    \texttt{hb711@georgetown.edu}
    \and
    Anat Bremler-Barr \\
    Tel Aviv University \\
    \texttt{anatbr@tauex.tau.ac.il}\\
}


%


\maketitle
\begin{abstract}

We present a novel yet simple and comprehensive DNS cache \systemFull~(\system), designed to integrate as a module in Intrusion Prevention Systems (IPS). \system{} addresses statistical DNS poisoning attacks, including those documented from 2002 to the present, and offers robust protection against similar future threats. It consists of two main components: a detection module that employs three simple rules, and a mitigation module that leverages the TC flag in the DNS header to enhance security.
Once activated, the mitigation module has zero false positives or negatives, correcting any such errors on the side of the detection module.

We first analyze \system{} against historical DNS services and attacks, showing that it would have mitigated all network-based statistical poisoning attacks, yielding a success rate of only 
0.0076\%
for the adversary. We then simulate \system{} on traffic benchmarks (PCAPs) incorporating  current potential network-based statistical poisoning attacks, and benign PCAPs; the simulated attacks still succeed with a probability of 
0.0076\%. This occurs because five malicious packets go through before \system{} detects the attack and activates the mitigation module. 
In addition, \system{} completes its task using only 20\%–50\% of the time required by other tools (e.g., Suricata or Snort), and after examining just 5\%–10\% as many packets. Furthermore, it successfully identifies DNS cache poisoning attacks—such as fragmentation attacks—that both Suricata and Snort fail to detect, underscoring its superiority in providing comprehensive DNS protection.

\end{abstract}

\section{Introduction}
\label{intro}

One of the cornerstones of our digital world is the Domain Name System (DNS)~\cite{mockapetris1987domain,mockapetris1987rfc1035}, which translates human-readable domain names into IP addresses.
As the Internet’s reliance on DNS grows, so does the complexity of this essential service.
Continual extensions, refinements~\cite{rose2012dname,andrews1998negative}, and security and privacy enhancements~\cite{arends2005protocol,bortzmeyer2016dns} add to this ever-increasing complexity, demanding a structured, comprehensive approach to effectively address emerging challenges.

In a DNS cache poisoning attack, an attacker injects false information into a DNS resolver’s cache.
This manipulation causes the resolver to return the wrong IP address for a target domain name, redirecting users to malicious websites without their knowledge. The attack can serve as a gateway to phishing \cite{chau2018adaptive}, credential theft \cite{roar_lk_domain_registry_2024}, malware infections \cite{securelist_dns_poisoning_2024}, and other online threats.
DNS cache poisoning remains one of the most persistent DNS-based attacks \cite{IEEE_Panix_Attack,kaminsky2009dns,securelist_dns_poisoning_2024,roar_lk_domain_registry_2024}, despite multiple attempts at mitigation \cite{schuba1993addressing,kaminsky2008black}.

\ignore{ In a cache poisoning attack~\cite{Akamai_DNS_Cache_Poisoning}, an attacker forges a response to a resolver query and delivers it to the resolver before the authentic response arrives. Thus, associating an IP address of its wish with a specific domain name that the attacker tricked the resolver to query about. 
Hence, re-directing users of this resolver to a malicious web site instead of the intended site. }
\ignore{
Within this landscape lies the persistent threat of DNS cache poisoning attacks~\cite{schuba1993addressing,kaminsky2008black}. A DNS cache poisoning attack is when an attacker inserts false information into a DNS server's cache, redirecting users to malicious sites. Users who are redirected by these false mapping can be lured to download viruses that can harm their devices~\cite{securelist_dns_poisoning_2024}, massive leakage of user credentials~\cite{roar_lk_domain_registry_2024}, etc. 
}


There are two primary types of DNS poisoning attacks.
The first and more popular method involves delivering spoofed authoritative responses to a resolver by faking the IP address of an authoritative name server, and guessing parameters such as the DNS transaction ID and source port to match a corresponding query issued by the resolver.
The second uses advanced methods—like BGP hijacking~\cite{morillo2021rov++} or MITM attacks~\cite{conti2016survey}—to bypass these guesswork steps, but is more complex to execute and therefore less common. Our study focuses on the former, more prevalent type of attack.

Proposed in 1997, the DNS Security Extensions (DNSSEC) protocol was designed to eliminate DNS cache poisoning attacks through cryptographic authentication. However, with limited global adoption (<30\%)\footnote{\url{https://www.myrasecurity.com/en/knowledge-hub/dnssec/}} and dependence on authoritative server implementation, DNSSEC has yet to achieve universal protection. Partial mitigations, such as increased randomness \cite{kaminsky2008black} and machine learning \cite{berger2021wrinkle}, address earlier attack vectors but prove less effective against more advanced threats \cite{li2023maginot,herzberg2012security}.

In the absence of a universal solution, DNS vulnerabilities are addressed reactively: when a new DNS cache poisoning attack is discovered, it usually receives a Common Vulnerabilities and Exposures (CVE) identifier, prompting vendors to release updates.
This reactive cycle requires ongoing vigilance and frequent patches to maintain security.

In this paper we address the challenges posed by DNS cache poisoning attacks, with a novel mitigation system called \systemFull~(\system). 
The system consists of two components: a detection module (Section~\ref{sub:rules}) and a mitigation module (Section~\ref{sub:mit_tr}).
The detection module applies three simple rules, derived from a comprehensive analysis of DNS cache poisoning attacks throughout history, effectively capturing the unique attributes of statistical DNS cache poisoning. 
The mitigation module uses the TC flag in the DNS header to force suspicious request/response transactions to switch from UDP to TCP—a more secure, connection-oriented protocol whose stateful handshake and sequence number management prevent spoofing (in addition to unauthorized data injections, and session hijacking).
An earlier deployment of a prototype of \system,  could have blocked with high likelihood all the attack vectors explored in this study, including CVEs that were discovered from 2012 until 2024.
(Employing the rules that are based only on the 2002 attack, could have prevented 60\% of subsequent DNS cache poisoning vector attacks up to 2024.)
\Harel{Two important practical properties that follow from our two‐stage system are;
(1) ensuring that suspected DNS cache poisoning packets are verified through a TCP session in the mitigation module, effectively eliminating possible false positives from the first stage; and
(2) allowing the detection module to err in the false‐positive direction to minimize false negatives, which decreases the attacker’s success rate to less than 1\%. 
Our system mitigates attacks and vulnerabilities described in academic literature and in DNS cache poisoning-related CVEs (e.g.,~\cite{li2023tudoor,jeitner2021injection}).}

The approach of building a system to prevent new attack vectors is also evident in the industry.
For example, Akamai reported that their retrospective analysis enabled their defense systems to block future attack vectors, although they did not provide specific details about the solution~\cite{akamai2024keytrap}. 

In summary, the contributions of this work are:
\begin{itemize}
\item \textbf{Quick and accurate detection with Minimal Overhead.}
\ignore{
Our system employs three detection rules that effectively capture  DNS cache poisoning attacks throughout history with
exceptional speed and precision.
}
Detecting poisoning attacks in under 1 second, after observing the first five packets.
\item \textbf{Mitigation with Zero False Positives.}
\system{} leverages the TC bit to switch suspected responses from UDP to TCP, blocking only responses that indeed masquerade the legitimate authoritative server.
\end{itemize}

The remainder of this paper is organized as follows. Background for this work is provided in Section~\ref{sec:back}, including the process of DNS resolution, the attacker's threat model, and the types of attacks examined in our work. \Harel{Section~\ref{sec:related} presents the historical DNS cache poisoning attacks and mitigation methods as related work.} Our methodology is given in Section~\ref{sec:meth}, including the detection and mitigation modules, their theoretical analysis, and implementation. The metrics used to assess our system are also presented in this section. The experiments are described in Section~\ref{sec:exp}. The results are presented in Section~\ref{sec:results}. A list of the CVEs that our system would have made superfluous is provided in Section~\ref{sec:cves}. A comparison of our system to Suricata and Snort is presented in Section~\ref{sec:tools}. A discussion on our system can be found in Section~\ref{sec:diss}. We conclude our work and suggest future work in Section~\ref{sec:conc}. Ethical considerations are elaborated in Section~\ref{sec:eth}, and compliance with open science policy is provided in Section~\ref{sec:open_sc}.\section{Background}
\label{sec:back}
\ignore{
This section provides an overview of the DNS resolution process (Section~\ref{dns_resolution}), followed by an illustration of the threat model used in DNS cache poisoning attacks (Section~\ref{threat_models}). We then discuss significant DNS cache poisoning attack vectors throughout history (Section~\ref{history}). Although there is insufficient evidence of widespread exploitation of these attacks in the wild, they exposed critical vulnerabilities in DNS resolvers, prompting immediate remediation efforts. An overview of these attack vectors is presented in Table~\ref{tab:hist_tab}.
}

\subsection{DNS resolution and poisoning attacks}
\label{dns_resolution}
When a DNS resolver receives a client query for domain $D$ it first checks if $D$ is in its cache; if not, it initiates a resolution process querying different name servers (NS) in the hierarchical DNS system, until it reaches the authoritative NS which holds the authority for answering queries about $D$.
Then, the authoritative NS sends a response with the IP address associated with $D$, back to the querying resolver. 
Upon receiving the response, the resolver validates that the response UDP packet has the correct source port, transaction ID (TXID), domain name ($D$), and that the source IP address is the IP address of the authoritative NS.
If validation succeeds the resource record (RR) in the response packet is inserted into the cache, with the corresponding time-to-live (TTL) value, and the corresponding IP address (A or AAAA type part of the record)  is delivered to the client.


In a cache poisoning attack an attacker inserts into the resolver cache an incorrect mapping of a domain name to an IP address of a malicious server (e.g., mimicking central bank server).
In the main variant of network based poisoning attacks this is done by tricking the resolver to accept a fraudulent response as if it came from the legitimate authoritative NS with the intended malicious mapping.  Mounting such an attack for domain $D$ involves variations on these  steps:
\begin{enumerate}
\setlength{\itemsep}{-2pt}
\item \vspace{-8pt}
(optional) Remove a record for $D$ from the resolver cache, if such exists.  This new step introduced here to strengthen the attack. It can be achieved with the CachFlush technique of \cite{afek2024flushing}.\\
\item \vspace{-12pt}
Trick the resolver to  query the authoritative NS of $D$, expecting a response with the IP address of $D$.
\Harel{This is achieved by a malicious client querying the resolver for the IP address of $D$, which is not in its cache due to step 1, or other methods, such as querying a fake subdomain of $D$ which is not in the cache~\cite{kaminsky2008black}.}\\
\item \vspace{-12pt}
While the resolver is expecting a response from the authoritative NS, provide it with a forged response with the source IP of the authoritative NS.\\
\item \vspace{-12pt}
For the forged response in step 3 to be accepted by the resolver it must have the correct port-number, TXID, and domain $D$, in addition to the correct IP address of authoritative NS. The port number and TXID, which are randomly generated by the resolver, are not known to the attacker who has to guess them.
\\
\item \vspace{-12pt}
\Harel{If the attacker's forged response is accepted by the resolver, it is injected to the cache, saving fake mapping of $D$ in the cache (directly, or via a glue record~\cite{kaminsky2008black}).}
\end{enumerate}

We distinguish between four major types of statistical poisoning attack; \(S\), \(S_{\text{\textit{Frag}}}\),
 \(B_{\text{\textit{Frag}}}\), \(S_{\text{\textit{OoB}}}\), some of the type may vriants on the steps:
\label{attack_types}

\begin{figure}[h!]
    \centering
    \includegraphics[width=\columnwidth]{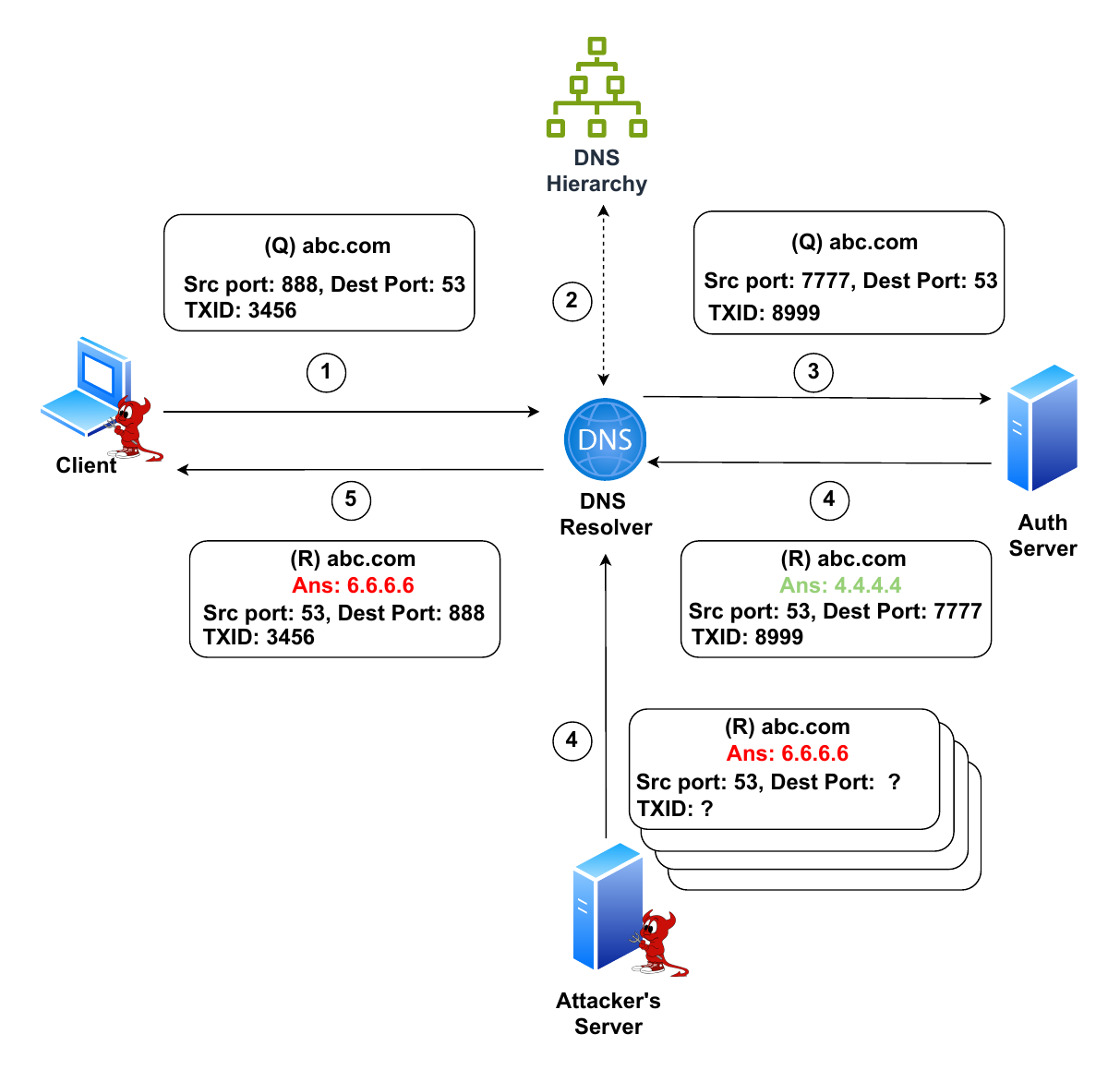} 
    \caption{\textbf{Simple DNS Cache Poisoning Attack} In a typical DNS cache poisoning attack (type $S$), the following steps occur. (A) A client controlled by the attacker queries a DNS resolver for a domain such as abc.com. (2) If the resolver does not have the response in its cache, it follows the DNS hierarchy to resolve the query. (3) At the bottom of the hierarchy, the resolver queries abc.com's authoritative server. (4) In parallel, both the correct response from the authoritative server and a set of fake responses from the attacker reach the resolver. The attacker attempts to guess necessary parameters—such as the transaction ID (TXID) or the source port of the resolver—to successfully poison the cache. It sends multiple responses based on the parameters it needs to guess. (5) If one of the attacker's packets has the correct parameters and arrives before the legitimate response, the resolver caches the fake response and returns it to the client.}
    \label{fig:attack_base}
\end{figure}

In type $S$ (see Figure~\ref{fig:attack_base}) in step 4 above, the attacker is trying many combinations of the (source) port number and TXID in the hope that one of them will have the correct number.
If neither of the two may be obtained then it has $\frac{1}{2^{32}}$ possibilities, and if one of two may be acquired by the attacker then there are still $\frac{1}{2^{16}}$ possibilities. 
The attacker strategy is to send many combinations in the hope to hit the correct one in order to succeed in the poisoning.
Each forged response guesses correctly either the TXID  or the port number with probability (if nothing else is known) $\frac{1}{2^{16}}$ or both with probability $\frac{1}{2^{32}}$
\footnote{
However, in the past, the attacker could have used the birthday attack \cite{flajolet1992birthday} in which it tricks the resolver into sending several requests concurrently for the target domain to the authoritative NS, then the probability of success in guessing a matching response is significantly increased to $\frac{1}{700}$ to $\frac{1}{400}$ due to the birthday paradox.  
This birthday paradox attack has been patched since 2003 by all major DNS vendors and in any event with our mitigation, its success would have been reduced to roughly $\frac{1}{500}$.}.

In type {\bf \(S_{\text{\textit{Frag}}}\)}, fragmentation is used as follows \cite{FragDNS2022}:
\Harel{The attacker first runs step 4, in which it sends a forged $2nd$ fragment containing the mapping it wants to be cached. Then, it follows step 2 by tricking the resolver to  query authoritative NS with a query whose response size does not fit into one UDP packet thus requiring fragmentation.}
While fragmented packets do not carry the port number nor the TXID, except for the first fragment, they do carry a fragmentation ID (IPID) a $16$ bit number. When the first fragment from the authoritative NS arrives at the resolver, it appends the already existing $2nd$ fragment if their IPIDs match. The response thus constructed from the two fragments is delivered to the resolver which then inserts  the poisonous mapping of the $2nd$ fragment into the cache. In this variant the attacker thus needs to send $2^{16}$ different IPIDs $2nd$ fragment in order to succeed.

In type {\bf \(B_{\text{\textit{Frag}}}\)}, the same procedure as with \(S_{\text{\textit{Frag}}}\) is followed except that it is a bullseye (B)  attacker, which knows the IPID number by other means and no guessing is required.  

In type {\bf \(S_{\text{\textit{OoB}}}\)}, the attacker exploits DNS resolvers that improperly handle DNS records in a response that pertain to domains outside the authority of the domain that was queried —a situation referred to as "out-of-bailiwick (OoB)."
In this attack, in step 4, the attacker includes in the forged response an OoB record with the mapping it wishes to insert into the cache \Harel{(e.g., a mapping for bank.com while the original query was for example.net. The .com domain is out of bailiwick for .net)}.
Due to inadequate validation, the resolver may erroneously cache the additional OoB mappings.
Consequently, future queries for subdomains of example.com may return malicious IP addresses under the attacker's control.

 

\subsection{Threat Model}
\label{threat_models}

This paper considers the threat of off-path network based DNS cache poisoning attacks, targeting resolvers during the DNS resolution process.
\ignore{
Given that a mapping of a domain name to an IP address is cached in the resolver, an attacker cannot corrupt this mapping until the TTL value expires\footnote{\Harel{A recent paper~\cite{afek2024flushing} found that the cache entries can be easily maliciously flushed from the cache.
This allow more sophisticated cache-poisoning attacks, which are also prevented by \system{}. }}.
}
We consider attackers mounting attacks of types 
 \(S\), \(S_{\text{\textit{Frag}}}\), \(B_{\text{\textit{Frag}}}\), \(S_{\text{\textit{OoB}}}\), as described above in \ref{dns_resolution}.
The resolver employs randomization techniques such as using random source ports/DNS  TXID\footnote{Another existing security measurement is randomizing the capitalization of letters in the domain name~\cite{dagon2008increased} using the 0x20 bit, which is validated by the resolver if employed. However, there is no sufficient evidence for 0x20 bit global utility. Also, for short domain names, its effectiveness is limited~\cite{sophos_dns_security}.} to prevent an attacker from guessing these parameters.
To counteract this approach, the attacker uses the techniques described above in \ref{dns_resolution}.

\ignore{
a client that initiates queries to the resolver\footnote{\Harel{In the past, birthday attacks were carried out using multiple queries to the same domain, which increased an attacker’s probability of successfully guessing the correct parameters~\cite{CERT-VULS-457875,CVE-2002-2211,li2024dnsbomb}.
To counteract this, vendors were advised to employ a technique known as \textit{query aggregation}. This method causes the resolver to generate a single resolution process for multiple queries from different clients requesting the same domain name. As a result, major} \Harel{vendors have implemented query aggregation~\cite{li2024dnsbomb}.}}.
}

Our threat model assumes the attacker brute-force predicts the DNS TXID and source port, and also the resolver lacks additional security measures that could prevent such attacks (e.g., DNSSEC).
Also, the attacker is assumed to use the UDP transport protocol, except if forced otherwise by~\system.

Attacks involving man-in-the-middle (MITM) positions, BGP hijacking (e.g.,\cite{dai2021ip,ballani2007study,birge2018bamboozling}), or direct compromise of the resolver are out of scope for this study.
we do not specifically consider DNS cache poisoning attacks on clients, e.g., forwarders, (e.g.,\cite{alharbi2019collaborative,ma2020intelligent}), which generally have less impact, as they target a single client. \Harel{Moreover, attacks on the DNS client can be mitigated by using our system with slight modifications. For example, embedding a syn-cookie in~\system's~TCP connections~\cite{afek_protecting_2002,afek2017network} can prevent the acceptance of the attacker's fake responses by the client.}  
Additionally, we exclude attacks that focus on improper processing of DNS responses~\cite{jeitner2021injection}. These attacks exploit syntactic vulnerabilities~\cite{CWE-89} and therefore require different mitigation strategies~\cite{elshazly2014survey}, which are beyond the scope of this paper.

\section{Related Work}
\label{sec:related}

\subsection{DNS Cache Poisoning Attacks Evolution}
\label{history}

In this section, we present the DNS cache poisoning attacks from the academic literature (see Table~\ref{tab:hist_tab}). \Harel{For brevity, we mention the type of the attacks in parentheses (e.g., 2008~\cite{kaminsky2008black} ($S$))}.
\begin{table}[!h]
    \centering
    \begin{tabular}{lcc||c}
        \toprule
        \textbf{Paper} & \textbf{Type} & \textbf{\#Pkts} & \textbf{\system} \\
        \midrule
        Schuba et al.~\cite{schuba1993addressing} \hfill 1993 & - & 1 & - \\
        V. Sacramento ~\cite{sacramento2002vulnerability} \hfill 2002 & $S$ & $2^{16}$ & R$\ell$1 \\
        Klein, Amit ~\cite{klein2007bind} \hfill 2007 & $S$ & $>$100 & R$\ell$1 \\
        Kaminsky, Dan ~\cite{kaminsky2008black} \hfill 2008 & $S$ & 200 * \textit{q} & R$\ell$1 \\
        Herzberg et al. ~\cite{herzberg2012security} \hfill 2012& $S$ & $2^{16}$ & R$\ell$1 \\
        Herzberg et al.~\cite{herzberg2012security} \hfill 2012 & $S_{Frag}$ & $2^{16}$ & R$\ell$2 \\
        Herzberg et al.~\cite{herzberg2013fragmentation} \hfill 2013 & $B_{Frag}$ & 1 & R$\ell$2 \\
        Herzberg et al.~\cite{herzberg2013vulnerable} \hfill 2013 & $S_{Frag}$ & $\sim2^{11}$ & R$\ell$2 \\
        Herzberg et al.~\cite{herzberg2013socket} \hfill 2013 &  $S, S_{Frag}$ & $2^{16}$ & R$\ell$1 \\
        Zheng et al.~\cite{zheng2020poison} \hfill 2020 & $B_{Frag}$ & 1 & R$\ell$2 \\
        Man et al. ~\cite{man2020dns} \hfill 2020 & $S$ & $2^{16}$ & R$\ell$1 \\
        Dai et al.~\cite{dai2021dns} \hfill 2021 & $S_{Frag}$ & 64 & R$\ell$1 \\
        Klein et al.~\cite{klein2021cross} \hfill 2021 & $S$ & $2^{16}$ & R$\ell$1 \\
        Jeitner et al.~\cite{jeitner2022xdri} \hfill 2022 & $S$ & $2^{16}$ & R$\ell$1 \\
        Jeitner et al.~\cite{jeitner2022xdri} \hfill 2022 & $S$ & $2^{16}$ & R$\ell$1 \\
        Li et al.~\cite{li2023maginot} \hfill 2023 & $S_{OoB}$ & $2^{16}$ & R$\ell$3 \\
        Heftring et al.~\cite{heftrig2023poster} \hfill 2023 & $S_{Frag}$ & $2^{16}$ & R$\ell$2 \\
        Li et al.~\cite{li2023tudoor} \hfill 2024 & $S$ & $2^{16}$ & R$\ell$1 \\
        \bottomrule
    \end{tabular}
    \caption{\textbf{Significant DNS Cache Poisoning Attacks.} The 'Type' column describes the type of the attack. Two major types are described - Bullseye ($B$) and Statistical ($S$). $B$ refers to an attack in which the targeted resolver sees only one packet the attacker generates. $S$ denotes an attack in which the attacker sends multiple packets to the resolver. 
    A sole $S$ refers to an attack that involves TXID/Port guessing, $S_{Frag}$ denotes a fragmentation attack using multiple packets, $B_{Frag}$ refers to a fragmentation attack using a single packet, and $S_{OoB}$ \Harel{refers to "Out-of-Bailiwick", the creation of packets that violate the Bailiwick rule}. The number of packets used for the attack is described in the $\#Pkts$ column. The $q$ value refers to the number of queries during the attack. The last column presents the~\system~rules discussed in Section~\ref{sec:meth}.}
    \label{tab:hist_tab}
\end{table}
The first documented DNS cache poisoning attack dates back to 1993~\cite{schuba1993addressing}, and involved creating a fake DNS response that arrives before the legitimate response.
However, since this attack is performed within the resolver itself, and our threat model assumes an attacker residing outside the resolver, we did not consider its implications in our study.
Subsequently, two $S$ type attacks were documented by Sacramento~\cite{sacramento2002vulnerability} and Klein~\cite{klein2007bind}. 
Both attacks assumed that the attacker knows the source port of the resolver because it is constant or can be determined from other sources. 
In both the attacker uses brute-force guessing on the TXID and sends fake responses with each possible TXID to the resolver in parallel with the legitimate response. 
At the time the attacks were published, the security measures included only semi-randomization of the Transaction ID (TXID). 
Therefore, the attacker only needed to guess the 16-bit TXID correctly. 

The next attack to consider is Dan Kaminsky's renowned DNS cache poisoning attack from 2008~\cite{kaminsky2008black} ($S$). In this attack, the attacker assumes that the resolver's source port is known and aims to poison the cache with malicious information. To achieve this, and bypass the limitation that the victim domain can be poisoned only when the TTL of the domain is expired,  the attacker generates a large volume of queries to non-existent subdomains of the target domain. For example, if the target domain is \textit{abc.com}, the attacker queries the resolver for \textit{1.abc.com}, \textit{2.abc.com}, and so on. For each query, the attacker guesses 200 TXIDs and sends fake responses to the resolver using these guessed TXIDs. Each forged response includes glue records for the authoritative name server of the target domain (e.g., \textit{ns.abc.com}) that map this name server to a malicious IP address controlled by the attacker.

Following Kaminsky's attack, a new countermeasure was implemented that involved randomizing both the transaction ID (TXID) and the source port of resolver queries, requiring an attacker to guess 32 bits to successfully spoof a response. Subsequent attacks aimed to reduce this guessing requirement back to 16 bits by eliminating the need to guess one of these values. The first notable attempt to address this challenge was by Hertzberg et al.\cite{herzberg2012security} ($S$), who focused on scenarios where the resolver is behind a Network Address Translation (NAT). They developed methods to pin or predict the NAT port, thereby reducing the need to predict both the TXID and the source port. In a subsequent study, Hertzberg\cite{herzberg2013socket} ($S$, $S_{Frag}$) described attacks where the name server's port is calculated using a query-generating zombie host. The attacker then sends manipulated responses, needing to guess only 16 bits. Man et al.\cite{man2020dns} ($S$) proposed an attack involving a Distributed Denial of Service (DDoS) on the authoritative server, combined with a UDP/ICMP port scan to uncover the resolver's source port mechanism, followed by sending fake responses with different TXIDs. Klein et al.\cite{klein2021cross} ($S$) demonstrated how an attacker could determine the resolver's source port by analyzing the state of the pseudo-random number generator, then only guessing the TXID to run a cache poisoning attack. Jeitner et al.\cite{jeitner2022xdri} ($S$) identified weaknesses in routers with built-in DNS resolvers, including predictable ID generation and constant source ports, which attackers could exploit by varying TXIDs or source ports. In the Tudoor attack~\cite{li2023tudoor} ($S$), the attacker first sends a benign query to the resolver. Subsequently, the attacker sends multiple DNS queries embedding port numbers within the domain name to identify the correct source port of the resolver. After determining the correct port, the attacker only needs to guess the TXID of the original benign query by sending multiple fake responses with different TXIDs.

An alternative approach to DNS cache poisoning leverages IP fragmentation to bypass the need for guessing the Transaction ID (TXID) or source ports. Hertzberg et al.\ \cite{herzberg2013fragmentation} ($B_{Frag}$) pioneered this method by analyzing how resolvers generate IP identification (IPID) values; by predicting the IPID sequence, an attacker can send a single malicious second fragment that the resolver mistakenly reassembles with a legitimate first fragment, leading to cache poisoning. In subsequent work, Hertzberg et al.\ \cite{herzberg2013vulnerable} ($S_{Frag}$) introduced a technique involving multiple second fragments with different IPIDs, allowing the attacker to manipulate the resolver into reassembling a legitimate first fragment with a fake second fragment by increasing the chances of IPID collision. Zheng et al.~\cite{zheng2020poison} ($B_{Frag}$) expanded on fragmentation attacks by involving both a forwarder \Harel{(a DNS server that forwards DNS queries to external servers for resolution)} and a resolver: in their attack, the attacker sends a malicious second fragment to the forwarder, which then reassembles it with a legitimate first fragment before forwarding it to the resolver. Dai et al.\ \cite{dai2021dns} ($S_{Frag}$) demonstrated that an attacker can induce fragmentation using ICMP packets and then send a malicious second fragment directly to the resolver; when the resolver processes a legitimate query, it reassembles the response using the attacker's fragment, resulting in cache poisoning. Heftrig et al.\ \cite{heftrig2023poster} ($S_{Frag}$) described an attack targeting DNSSEC responses. By first triggering fragmentation and then sending a counterfeit DNSSEC second fragment, the attacker causes the resolver to include the fake fragment in the reassembled response from the authoritative server, thus poisoning the cache even in the presence of DNSSEC.

The last approach was demonstrated by Li et al.~\cite{li2023maginot} ($S_{OoB}$), in which an attacker induces a CDNS \Harel{(a DNS server that analyzes incoming queries and conditionally forwards them to designated resolvers or forwarders for tailored responses based on its policy)} to query its own domain. The attacker then sends a spoofed response, which includes an NS record that delegates a top-level domain (TLD) to the attacker's authoritative server—a clear violation of the "out of bailiwick" rule. As a result, the CDNS caches this incorrect delegation to the attacker's server.

\subsection{Related Poisoning Prevention Systems}

Throughout the years, several efforts have focused on hardening the DNS protocol against attacks. Probably the most famous one is DNSSEC~\cite{arends2005dns}, which started back in 1997. This defense allows for the cryptographic authentication of DNS records, thus preventing DNS cache poisoning attacks via counterfeit packets. DANE~\cite{barnes2011use,gudmundsson2014adding,zhu2015measuring} utilizes the structure of DNSSEC to authenticate domains using TLS certificates. The adoption of these solutions is limited by the global implementation of DNSSEC in resolvers, which is less than 30\%\footnote{https://www.myrasecurity.com/en/knowledge-hub/dnssec/}. DNSCurve~\cite{dempsky2010dnscurve}, introduced in 2009, attempted to encrypt DNS traffic to prevent interception of queries and responses, but it was not widely adopted. Since then, other approaches have tried to make DNS more resilient to privacy and security issues~\cite{bortzmeyer2016dns,hu2016specification,reddy2017dns,hoffman2018dns}. DNS over HTTPS\footnote{HTTPS itself may prevent users from accessing a malicious website without a valid certificate. However, as noted by Doe~\cite{Doe2023}, DoH does not offer this safeguard. Furthermore, the resolver—which is the focus of this paper—is not secured when handling HTTPS traffic.} (DoH)~\cite{hoffman2018dns} allows DNS queries and responses to be sent over secure HTTPS traffic and has been implemented since 2018 by services from Google, Amazon, and Cloudflare. Although it allows safer transit of DNS queries, there is no transparency in how DNS queries are processed on the sender's side\footnote{https://www.csoonline.com/article/575131/the-status-quo-for-dns-security-isn-t-working.html}. \Harel{Although DOH can prevent attacks on the traffic between clients and resolvers, it does not prevent the poisoning of the resolver's cache. Thus, DOH does not prevent the attacks examined in our work.}

Other works have abstracted the DNS protocol to detect anomalous or malicious DNS traffic. One of the earliest efforts in this direction assessed DNS cache poisoning attacks~\cite{alexiou2010formal}, focusing solely on Kaminsky's attack using the probabilistic model checker PRISM~\cite{kwiatkowska2002prism}. A similar approach was presented in~\cite{deshpande2013formal}. The SPIN model checker~\cite{holzmann1997model} was utilized in another study with finite state machine models in~\cite{zhanga2016model} to detect abstracted (but non-specific) DNS cache poisoning attacks.

A DNS cache poisoning mitigation was devised by Laraba et al.\cite{laraba2022detecting} using Petri nets\cite{peterson1977petri} and P4 programmable switches. However, it addressed only simple DNS cache poisoning, similar to Kaminsky's attack~\cite{kaminsky2008black}.

Other approaches have employed machine learning algorithms to detect DNS attacks. For instance, Anax~\cite{antonakakis2010centralized} used \Harel{a machine learning (ML)} approach based on heuristics from 300,000 servers around the world to detect cache poisoning attacks. More details can be found in~\cite{jin2019detection}.

All the above-mentioned works analyzed the attacks through heuristics, patterns, and machine-learning ideas to tackle different DNS cache poisoning attacks. However, no solution tried to cover all DNS cache poisoning attacks at once. \Harel{In the current work, we suggest such a solution with a set of three simple rules and a mitigation technique.}
\section{Poisoning Attacks Detection and Mitigation}
\label{sec:meth}
\Harel{We present a DNS cache poisoning attack prevention system consisting of two modules: rule-based detection (Section~\ref{sub:rules}) and mitigation (Section~\ref{sub:mit_tr}).
Once the mitigation has kicked in the module is highly accurate, with zero false positives and zero false negatives, meaning it prevents all spoofed DNS responses from poisoning the cache and allows those coming from an authentic IP address (of the authoritative server) to be delivered as is to the resolver.  Furthermore, since the mitigation kicks in after 5 attacker packets, the probability of successful poisoning is 5 * the probability of each of the five packets guessing correctly both the port and TXID.  This is equal to 0.00152\%.
For efficiency, mitigation is applied selectively to responses flagged as suspicious by the detection module.}
\Yehuda{Although the detection module may occasionally err in the false positive direction, overall it does not result in blocking a legitimate response since these flagged responses are passed to the mitigation module, which ultimately blocks only the actual spoofed responses.
This property allows us to minimize the false negatives of the detection module and reach a close-to-perfect recall rate.
After describing the two modules,}
\Harel{we delve into the algorithmic approach of the detection module and present our complete mitigation system,~\system~ (Section~\ref{sub:r1}-\ref{sec:algo}).
Finally, we discuss the metrics used to evaluate our system (Section~\ref{sub:met}).}

\subsection{Detection Module}
\label{sub:rules}

\Harel{According to Table~\ref{tab:hist_tab}, the type of 60\% of the DNS attacks examined in our work involves guessing either the TXID or source ports (type $S$), $2^{16}$ possibilities for each.
An additional 28\% use the second fragment as an attack vector (type $S_{Frag}$/$B_{Frag}$).
The two exceptions, are Schuba et al. and Li et al.; the Li et al. attack exploits incorrect Bailiwick rule enforcement (type $S_{OoB}$), while the Schuba et al. attack, though listed, is outdated and not feasible today.}

Based on this trend, we devised three rules, each designed to detect a different type of potential DNS cache poisoning attack:

\begin{enumerate}
\item \textbf{Excessive Guessing of TXID/Port (R$\ell$1):} We monitor the number of DNS response packets that differ from each other only in one common field—the port or the TXID, within a small window of time.
\Yehuda{When this number crosses a threshold (a system parameter, e.g., $5$) for the same DNS query, we flag these responses as a potential poisoning attack and pass them to the mitigation module}.
\item \textbf{Fragmentation (R$\ell$2):} The second dominant behavior of DNS cache poisoning is sending fake second fragment packets in response to pre-requested DNS queries. \Harel{ When the first fragment of a packet is obtained, it is passed to the mitigation module. Any other fragments are discarded to prevent fragmentation-based attacks.} 

    \item \textbf{Out of Bailiwick (R$\ell$3):} \Harel{We refer here to the 'out of bailiwick' case, where the domain being queried falls outside the authority of the DNS server responding to the request. Packets that meet this rule are identified as potential attack packets as well and are passed to the mitigation module}.
\end{enumerate}
The last column of Table~\ref{tab:hist_tab} summarizes the matching of an attack and a rule, based on their DNS cache poisoning attack type.
\Harel{The efficient implementation of the rules is presented in Section~\ref{sec:algo}.}

\subsection{Mitigation Module}
\label{sub:mit_tr}

\Yehuda{When a response is suspected of being a poisoning its content is being erased and the corresponding query/response conversation is moved from DNS over UDP to DNS over TCP, by setting the TC flag to true on the response that is being forwarded to the target resolver. This way the resolver has a higher guarantee that it communicates with an authentic IP address due to the three-way handshake of TCP, thus mitigating poisoning that does not come from the intended name server itself.}
\Harel{
\ignore{
When a potential attack is detected, we defend the target resolver by switching to the DNS protocol.
We signal the resolver to transition from DNS-over-UDP to DNS-over-TCP by setting the TC flag in the DNS header to true.
}
}
\Yehuda{The TC flag~\cite{RFC1035} was originally designed to indicate that a packet was truncated due to the packet's length (i.e., the packet is larger than the maximum transmission unit (MTU)) and thus should be discarded by the resolver, and be requested again over TCP.
While previous works~\cite{afek2017network,vixiedns,afek_protecting_2002} used the TC flag to defend from various attacks, to the best of our knowledge, it has not yet been systematically applied to mitigate DNS cache poisoning attacks.
The TC flag was used randomly for clients suspected of being hijacked~\cite{vixiedns} or to mitigate spoofed DNS DDoS attacks \cite{afek_protecting_2002,afek2017network}.
In contrast, the TC bit is used here to systematically protect resolvers from DNS cache poisoning attacks.}

\Harel{
Practically, when a suspicious response is identified, its TC flag is set to true (TC=1). Additionally, all data of the answer, authoritative (auth), and additional (add) parts of the answer are removed from the response\footnote{\Harel{In the case of R$\ell$2, the fragment contains partial information. Since only the first fragment is sent to the mitigation module, we can derive the required field values from the question section within this fragment to generate the truncated response (e.g., qname, TXID, etc.). To the best of our knowledge, there is no documentation in the official RFCs~\cite{RFC1035, rfc6891} specifying which fields must be extracted from the truncated packet. We assume that the DNS layer in the truncated packet provides the necessary information to initiate a new TCP session with the authoritative server.}}.}
\Harel{A recent measurement of DNS resolvers by APNIC Labs \cite{apnic2024dns} used ad-based measurement techniques to investigate the use of information from truncated packets. By examining the queries of approximately 394 million users around the world (primarily querying their default ISP's resolvers, with estimated hundreds of thousands of unique resolvers), APNIC found that 0.05\% of these queries used information from truncated packets, even though such data should be disregarded. To mitigate the use of potentially compromised data, we erase any data from truncated packets before forwarding the packet to the intended resolver. Additionally, APNIC observed that 2.67\% of resolvers do not resend a TCP query to the authoritative server after receiving a DNS response with the TC flag set to true. For comparison, Moura et al. \cite{moura2021fragmentation} found a lower rate of 1.78\% among resolvers in the Netherlands. An alternative mitigation option that might have addressed the issue with these non-cooperating resolvers could have involved setting the TTL value to zero, signaling that the information in the packet immediately expires. However, more than 8\% of resolvers ignore this explicit expiration signal by extending the response's TTL~\cite{bhowmick2023ttl}, limiting the applicability of this approach. Thus, our system is not 100\% compatible with resolvers that do not initiate a TCP session when receiving a truncated packet.}

\Harel{The mitigation module may send multiple packets (attacker packets) with the TC flag set to true to the resolver. Because the resolver validates both the TXID and the source port, most of these packets will not match the original query and will be discarded. The resolver will accept only the packets that match both parameters. Once a matching packet is accepted, the resolver opens TCP sessions with the authoritative NS. Even if the attacker matches both parameters, the transition to TCP ensures that the resolution process is robust against cache poisoning (e.g., requiring the correct sequence number, window length).}

\subsection{R$\ell$1 Algorithmic Approach}
\label{sub:r1}
\Harel{In R$\ell$1, we detect a high volume of packets with identical fields, except for the port or TXID. From Table~\ref{tab:hist_tab}, we see that the typical packet count for the $S$ type is 65,535—reflecting all possible values for either the port or TXID\footnote{Historically, in attacks such as those by \cite{kaminsky2008black,sacramento2002vulnerability,klein2007bind}, the randomness of both TXID and port was minimal, allowing shorter exhaustive searches. More recent attacks (e.g.,~\cite{li2023tudoor,jeitner2022xdri}) circumvented guessing the TXID and source port but still required navigating through the full range of 65,535 possible TXID or port guesses.}. Therefore, in R$\ell$1, a method is essential to differentiate benign DNS query responses—which are usually limited to one or two packets—from malicious cases, where thousands of responses may appear for the same query. Our objective is to develop an efficient, resource-light solution with a low error rate.}

To address this, we review solutions presented in previous works. An interesting algorithmic approach was used by Afek et al.~\cite{afek2016efficient} and Nadler et al.~\cite{ozeryinformation} to identify anomalous behavior in networks using Heavy Hitters. Specifically, Afek et al.~\cite{afek2016efficient} used it for DNS DDOS attack detection, and Nadler et al.~\cite{ozeryinformation} for DNS tunneling attack detection. Both works utilized the constant size memory of heavy hitters to address large amounts of traffic. Another useful approach was demonstrated in DNSGuard~\cite{duan2024dnsguard}, an ML-based technique for mitigating DNS attacks, using Count-Min Sketch~\cite{cormode2009count} (CMS). 

\begin{figure*}[h!]
    \centering
    \begin{subfigure}{0.49\textwidth}
        \centering
        \includegraphics[width=\textwidth]{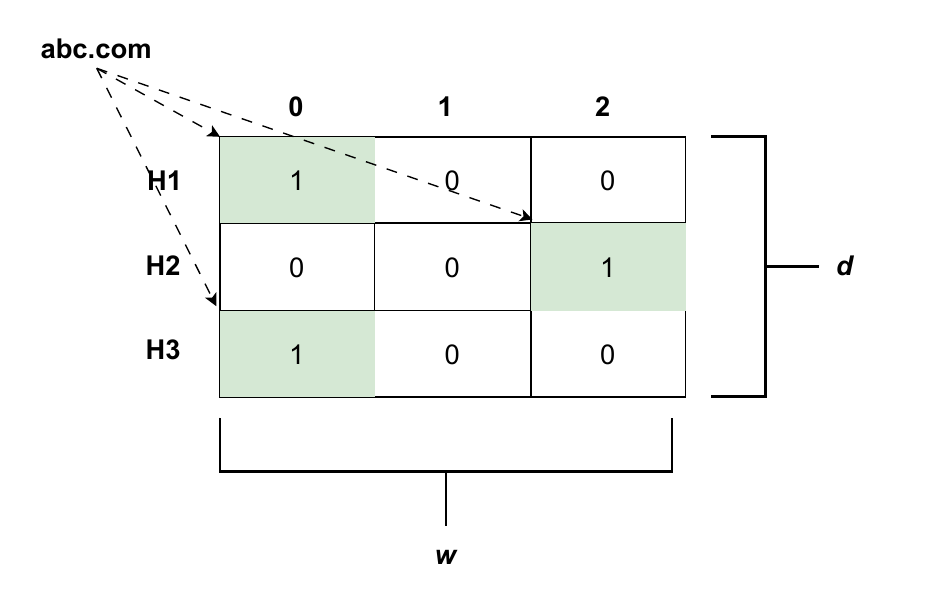} 
        \caption{}
        \label{fig:cms}
    \end{subfigure}
    \hfill
    \begin{subfigure}{0.4\textwidth}
        \centering
        \includegraphics[width=\textwidth]{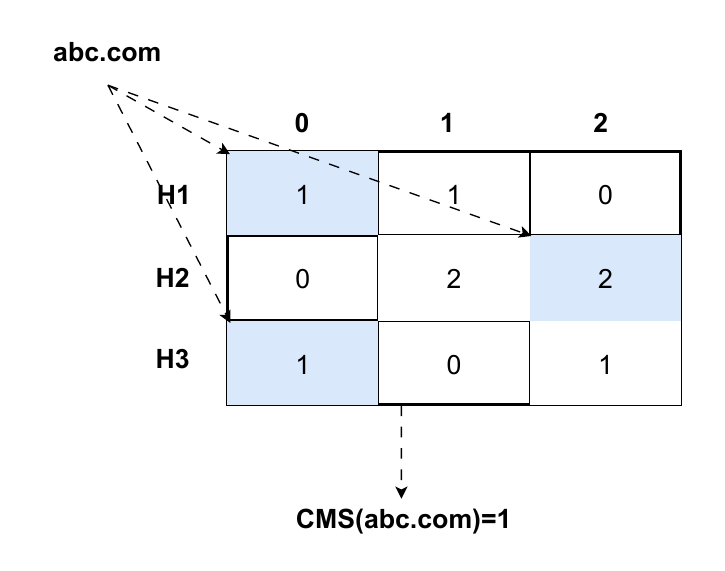} 
        \caption{}
        \label{fig:cms_query}
    \end{subfigure}
    \caption{\Harel{\textbf{CMS illustration.} CMS is defined by a number of hashed functions \(d\), and the amount of counters per hash function \(w\). In the subfigures, we describe each hash function as $H_i$.
    When a new element is obtained, it is hashed multiple times, and each hashed value is mapped to a cell in the respective row, increasing the counter in that cell (\subref{fig:cms}). When an element is estimated, it is hashed using each one of the hash functions, gathering all values from the appropriate cells, and obtaining the minimum value of these cells (\subref{fig:cms_query}).}}
    
    \label{fig:cms_combined}
\end{figure*}

Thus, we compare three different algorithms previously used for frequency estimation in DNS attacks: Count-Min Sketch (CMS), Fixed-size Distinct Weighted Sampling (dwsHH), and Fixed-threshold Distinct Weighted Sampling (WS).\\
    \textbf{Count-Min Sketch (CMS)} is a compact data structure used to estimate item frequencies in a data stream. It consists of a table with multiple rows, each linked to a unique hash function, and multiple columns representing counters. When an item arrives, it is hashed using each function, and the resulting hash values indicate which counters to increment. The item's frequency is estimated by taking the minimum value among these counters. CMS's memory usage depends on the desired error rate (\(\epsilon\)) and confidence level (\(1 - \delta\)), which determine the number of hash functions (depth, \(d\)) and counters per hash function (width, \(w\)). An illustration of CMS is presented in Fig.~\ref{fig:cms_combined}.

\textbf{Fixed-size Distinct Weighted Sampling for Heavy Hitters (dwsHH)} identifies frequently occurring items (heavy hitters) in a data stream using a fixed memory size. It maintains a weighted sample of items based on their frequency. When a new item arrives, the algorithm decides whether to add it to the sample, depending on its weight and the sample's current state. The memory usage is influenced by the cache size (\(k\)), the sampling window length (\(\ell\)), and the logarithm of the monitored item count (\(m\)). The sample dynamically updates, keeping the most significant items while minimizing memory use.

\textbf{Fixed-threshold Weighted Sampling (WS)} identifies significant or "heavy" items in a data stream using a fixed threshold (\(\tau\)). Items are sampled if their weight or frequency exceeds this threshold, which helps focus on important items without tracking every item in the stream. Upon the arrival of a new item, its weight is compared to \(\tau\); if it exceeds the threshold, it is added to the sample. This method optimizes memory and computation by focusing on items that significantly affect data distribution, reducing errors in frequency estimation for less significant items.

The comparison of the three algorithms is based on memory usage, error rates, and inference time for various numbers of domains (10, 100, 1000, and 10K). The full analysis can be found in Appendix~\ref{appendix_r1}. We describe here the comparison compactly, in Table~\ref{tab:comparison}. 

\begin{table}[h!]
\begin{tabular}{|c|p{1.5cm}|c|c|}
\hline
\textbf{Method} & \textbf{Memory } & \textbf{Error} & \textbf{Inference} \\ \hline
\textbf{CMS} & $w \times d \times \text{|counter|}$ & $\epsilon = \frac{e}{w}$ & $O(d)$ \\ 
\hline

\textbf{dwsHH} & $O(k\ell \log m)$ & $O\left(\frac{1}{k} \sum_y w_y + \frac{w_y}{\sqrt{2\ell}}\right)$ & $O(\log N)$ \\ 
\hline

\textbf{WS} & $O(\tau \sum_y w_y \cdot \ell \log m)$ & $O\left(\tau^{-1} + \frac{w_y}{\sqrt{2\ell}}\right)$ & $O(\tau N)$ \\ \hline
\end{tabular}
\caption{Comparison of memory usage (Memory), error rates (Error), and inference time (Inference) for CMS, dwsHH, and WS. Specific details are presented in Sections~\ref{sub:mem}-\ref{sub:infer}.}
\label{tab:comparison}
\end{table}

\subsubsection{Memory Usage Calculation}
\label{sub:mem}

\begin{enumerate}
    \item 
    \textbf{CMS}: The $w$ and $d$ values are fixed, and for optimum utility (will described in Section~\ref{sec:exp}), we picked $d=5,w=200$. The counter is given as an int (4 bits). The resulting memory usage is 4k bits.
    \item \textbf{dwsHH}:  Reasonable values used to compare are a cache size of \( k = 100 \), a sampling window size of \( \ell = 32 \), and a log of monitored size \( m \approx 0.1N \), where $N$ is the size of the domains' list.
    \item \textbf{WS}: the fixed threshold $\tau$ is 0.01. Other parameters are identical to the dwsHH.
\end{enumerate}

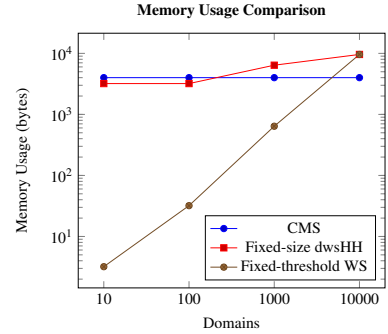
\begin{figure}[h!]
    \centering
    \resizebox{0.6\columnwidth}{!}{
    \begin{tikzpicture}
        \begin{axis}[
            title={\textbf{Memory Usage Comparison}},
            xlabel={Domains},
            ylabel={Memory Usage (bytes)},
            legend pos=south east,
            ymode=log,
            log basis y={10},
            xtick={10,100,1000,10000},
            xticklabels={10,100,1000,10000},
            xmode=log,
            log basis x={10}
        ]
        \addplot coordinates {(10, 4000) (100, 4000) (1000, 4000) (10000, 4000)};
        \addplot coordinates {(10, 3200) (100, 3200) (1000, 6400) (10000, 9600)};
        \addplot coordinates {(10, 3.2) (100, 32) (1000, 640) (10000, 9600)};
        \legend{CMS, Fixed-size dwsHH, Fixed-threshold WS}
        \end{axis}
    \end{tikzpicture}
    }
    \caption{Memory usage comparison, between CMS and Heavy Hitters. The CMS algorithm gets high memory usage but is constant. At its peak, the CMS is less memory-consumable than the heavy hitters.}
    \label{fig:memory}
\end{figure}

The comparison of memory usage is presented in Fig.~\ref{fig:memory}. It is shown that the memory usage of CMS is high but constant, and the heavy hitters' memory usage is increasing. With a high number of domains (10k), the CMS utilizes less memory.

\subsubsection{Error Rates Calculation}

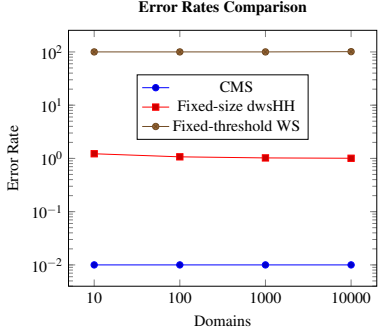
\begin{figure}[h!]
    \centering
    \resizebox{0.6\columnwidth}{!}{
    \begin{tikzpicture}
        \begin{axis}[
            title={\textbf{Error Rates Comparison}},
            xlabel={Domains},
            ylabel={Error Rate},
            legend style={
        at={(0.5,0.7)},  
        anchor=center   
    },
            ymode=log,
            log basis y={10},
            xtick={10,100,1000,10000},
            xticklabels={10,100,1000,10000},
            xmode=log,
            log basis x={10}
        ]
        \addplot coordinates {(10, 0.01) (100, 0.01) (1000, 0.01) (10000, 0.01)};
        \addplot coordinates {(10, 1.223) (100, 1.071) (1000, 1.022) (10000, 1.007)};
        \addplot coordinates {(10, 100.125) (100, 100.125) (1000, 100.125) (10000, 101.25)};
        \legend{CMS, Fixed-size dwsHH, Fixed-threshold WS}
        \end{axis}
    \end{tikzpicture}
    }
    \caption{Error rates comparison, between CMS and Heavy Hitters. The error rate of CMS is well under the heavy hitters, in order of magnitude ($\sim0.01$ vs more than 1, or more than 100).}
    \label{fig:error}
\end{figure}

All parameters for the three algorithms are the same as in the previous subsection.
The comparison of error rates is presented in Fig.~\ref{fig:error}. It can be seen that the error rate of CMS ($\sim$0.01) is less than the error rates of the heavy hitters (more than 1 and more than 100, respectively).

\subsubsection{Inference Time Calculation}
\label{sub:infer}

\begin{figure}[h!]
    \centering
    \resizebox{0.6\columnwidth}{!}{
    \begin{tikzpicture}
        \begin{axis}[
            title={\textbf{Inference Time Comparison}},
            xlabel={Domains},
            ylabel={Inference Time (operations)},
            legend pos=south east,
            ymode=log,
            log basis y={10},
            xtick={10,100,1000,10000},
            xticklabels={10,100,1000,10000},
            xmode=log,
            log basis x={10}
        ]
        \addplot coordinates {(10, 3) (100, 3) (1000, 3) (10000, 3)};
        \addplot coordinates {(10, 1) (100, 2) (1000, 3) (10000, 4)};
        \addplot coordinates {(10, 0.1) (100, 1) (1000, 10) (10000, 100)};
        \legend{CMS, Fixed-size dwsHH, Fixed-threshold WS}
        \end{axis}
    \end{tikzpicture}
    }
    \caption{Inference time comparison, between CMS and Heavy Hitters. CMS's inference time is constant at 5, where the other algorithms increase based on the number of domains.}
    \label{fig:inference}
\end{figure}
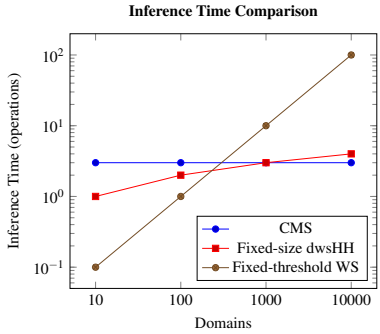

All parameters remain consistent with those in the previous subsections. As shown in Fig.~\ref{fig:inference}, the comparison of inference times indicates that the CMS query time stays constant—requiring only five hash function calculations—whereas the query time for the heavy hitters' methods increases with the number of domains.

Following the above analysis, we picked CMS for R$\ell$1. Compared to the other algorithms, CMS is less memory consumable with a high volume of domains, has a low error rate, and its reference time is constant.

\subsection{R$\ell$2 \& R$\ell$3 Algorithmic Approach}
\label{sub:r2}
\Harel{In R$\ell$2, fragments are identified as potential indicators for a DNS cache poisoning attack. Instead of storing data, we monitor the packet's offset and the MF (More Fragments) flag. When the offset equals zero and the MF flag is set (indicating the first fragment), fragmentation is detected.}
In R$\ell$3, we follow the Bailiwick rule. Any packet that does not enforce this rule, is identified as suspicious. Here as well, we do not need to store any data for future calculation.
We enforce both rules without any additional resource consumption.

\subsection{\system~algorithm}
\label{sec:algo}
Our system's final algorithm \Harel{(full pseudo code in Appendix~\ref{appendix_algo})} follows the match-action steps~\cite{mckeown2008openflow} for processing a given set of DNS response packets~$P$:

\begin{enumerate}
\item Get the CMS table, global counter, threshold, window size $W$, an initial time $t$ (first packet), a blank map $domainMap$, and a global boolean variable $action$.
    \item For each packet $p_i$ in $P$:
    \begin{enumerate}
        \item If applying R$\ell$1, R$\ell$2, or R$\ell$3 on $p_i$ matches, and $p_i$ is not null, set the action to "truncate". Erase the packet's answer/auth/add records and set the TC flag as true. 
        Else, the action is "forward". 
        \item Finally, send the packet to the destination.
        
    \end{enumerate}
    \item \textbf{R$\ell$1}(Algorithm~\ref{alg:Ru1}): 
    \begin{enumerate}
        \item Add the packet's domain name to the CMS table.
        \item Periodically, based on the counter value, check if the frequency of this domain in the CMS table is above the threshold. 
    \Harel{If so, check if this domain exists in $domainMap$ already. If so,} 
    \Harel{return true.} Elsewhere,
    if it is a new domain, save the domain in $domainMap$. Then, return true. 
    \item Increment the global counter.
    \item Periodically, based on the time value, check if we covered the whole window $W$. If so, empty the CMS, zero the counter, and set $t$ to the time of the current packet. 
    \item Return false.
    \end{enumerate}
    
    \item \textbf{R$\ell$2}(Algorithm~\ref{alg:Ru2}): If the offset of the packet equals zero and the MF flag is set to true, return true. Elsewhere, if the offset of the packet is greater than zero, the packet is nullified. Then, return false.
    \item \textbf{R$\ell$3:}(Algorithm~\ref{alg:Ru3}): If the answer, authoritative, or additional records of $p_i$ do not comply with the Bailiwick rule, return true. Elsewhere, return false.
\end{enumerate}

\subsection{Metrics}
\label{sub:met}
\Harel{We evaluated our system by measuring the attacker's success rate (ASR), a metric used to assess the effectiveness of DNS cache poisoning attempts. Given that common DNS cache poisoning attacks typically involve sending a high volume of packets, our goal was to determine the likelihood of an attacker's success. In this evaluation, we assume a 100\% success rate if the attacker transmits all intended packets without detection. However, if only a portion of the packets are sent before detection, the success rate is scaled according to the proportion successfully transmitted. For instance, if only 5 out of an intended 1000 packets are sent before detection, the success rate is calculated as 0.005.}

\Harel{We further evaluated our system's capability to accurately distinguish between benign and malicious packets by measuring the false positive (FP) rate, which indicates instances where benign packets are mistakenly classified as malicious.}

\Harel{In cases where the system identifies a suspicious packet, it triggers a truncated packet to the resolver to initiate a TCP connection with the authoritative name server. This mechanism ensures that the resolver opens a TCP connection to obtain the correct resolution, which is then passed on to the client. Only valid responses (matching a resolver's outgoing query) will be processed, the others will be dropped by the resolver. As a result, each client's query is answered reliably, even if an attacker attempts to poison the resolver. Thus, while false positives are analyzed, they do not actually impact the resolution process.}  

\section{Experiments}
\label{sec:exp}
In this section, we present the technical aspects of~\system~and describe how we evaluated its performance using a simulated environment. We implemented the rules and their testing framework in Golang and Python. The complete source code is publicly available in our GitHub repository~\footnote{https://anonymous.4open.science/r/Poisoning-Prevention-System-EE50}.
In the simulated environment, we collected and reconstructed a rich set of PCAP files containing different scenarios, including interleaved attack and benign packets, and legitimate DNS traffic to assess~\system's performance metrics.
This approach allowed us to test various configurations and conditions in a controlled and reproducible manner. 
All simulations were executed on a desktop machine with an 11th Gen Intel® Core™ i7-1185G7 @ 3.00GHz processor and 32 GB of RAM.
Using this setup, we conducted four distinct experiments to evaluate different facets of~\system:

\begin{enumerate}
    \item An experiment that mimics an extensive set of packets for testing R$\ell$1. To match the dominant attacks of the  $S$ model (as past attacks~\cite{kaminsky2008black,sacramento2002vulnerability,klein2007bind} have low success rates with today's defense settings), we generated a single query from the attacker to the resolver, 65,535 responses for this query with a different TXID, and an authentic response from the authoritative server.
Following the attack rate of approximately 400 ms as used in~\cite{li2023tudoor}, we timed the generation of the attack packets accordingly, aiming to send all guesses quickly enough to beat the correct response. 
     To test whether~\system~can identify attacks in the presence of benign traffic, we added 1,000 randomly generated noise packets—simulating valid response packets for other domains queried by other clients, with domains taken from the Tranco list~\cite{pochat2018tranco}—following the number of queries per second suggested by~\cite{li2024dnsbomb}. These noise packets simulate a more realistic network environment where benign traffic is sent to the resolver alongside the attacker's packets. Additionally, we explored different numbers of hash functions ($d=2,3,4,5$) and varying numbers of cells for each hash function ($w=100,200,500$), omitting lower $w$ values as they provided ineffective results.
    \item An experiment with an attack that mimics the 2nd fragmentation method. This experiment tested R$\ell$2.
    \item An experiment with an attack that violates the bailiwick rule. This experiment tested R$\ell$3.
    \item \Harel{The last experiment utilized benign data from Mendeley Data~\cite{singh2019}, which comprises over 45 million DNS packets collected over a period of 1.5 days from more than 4,000 users within an academic campus. This dataset, split into multiple files by hour as described in the original study, was analyzed by focusing solely on the DNS responses received by the resolver (IP address 172.31.1.6, as documented in~\cite{singh2019}) from servers in the DNS hierarchy. This focus is because responses from these servers represent the primary vector for potential DNS cache poisoning attacks. After filtering, approximately 6 million packets remained. This benign dataset served to evaluate the system's false positive (FP) rate, verifying that it does not incorrectly flag benign data as malicious at an excessive rate.}
    
\end{enumerate}

In all our experiments, we set the parameters to $\tau$=5 packets, $N$=1 packet, and a window $W$=1 second. These threshold and frequency-checking settings provide sufficient overhead for normal DNS traffic, typically involving 1–2 responses per query. The one-second window accommodates the most proficient attack reported by Li et al.~\cite{li2023tudoor} (approximately 400 ms) and simplifies the noise addition calculation. Therefore, we selected a one-second window for our experiments.

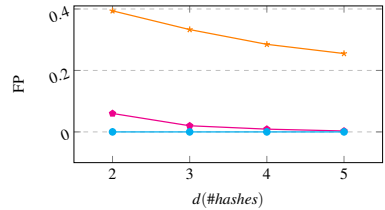
\begin{figure}[!h]
\resizebox{0.6\columnwidth}{!}{
\begin{tikzpicture}
    \begin{axis}[
        xlabel={\( d (\#hashes)\)},
        ylabel={FP},
        xmin=1.5, xmax=5.5,
        ymin=-0.1, ymax=0.41,
        xtick={2, 3, 4, 5},
        ytick={0, 0.2, 0.4},
        legend style={at={(1.1,1)},anchor=north west}, 
        y tick label style={rotate=20,anchor=east},
        legend columns=3,
        ymajorgrids=true,
        grid style=dashed,
        width=\columnwidth, 
        height=0.6\columnwidth, 
        legend to name=combinedlegend 
    ]

    \addplot[
        color=orange,
        mark=star,
        thick
    ]
    coordinates {
        (2, 0.394) (3, 0.333) (4, 0.285) (5, 0.255)
    };
    \addlegendentry{\( W = 100 \)}
    
    \addplot[
        color=magenta,
        mark=pentagon*,
        thick
    ]
    coordinates {
        (2, 0.060) (3, 0.020) (4, 0.009) (5, 0.003)
    };
    \addlegendentry{\( W = 200 \)}
    
    \addplot[
        color=cyan,
        mark=oplus*,
        thick
    ]
    coordinates {
        (2, 0.000) (3, 0.000) (4, 0.000) (5, 0.000)
    };
    \addlegendentry{\( W = 500 \)}
    
    \end{axis}
\end{tikzpicture}
}
\caption{\textbf{FP as a Function of $d$ Value - Interleaved Attack \& Benign Data.} Different $d$ values for CMS, and their FP rate for $w$ values. The attacker got a constant 0.0076\% success rate (5 packets), thus, it is not explicitly graphed. For FP rates, with $w=100$, the FP rate goes from 40\% with 2 hash functions ($d=2$), to 25\% with 5 hash functions ($d=5$). With $w=200$ and $d=5$, the FP rate is below $1\%$. For $w=500$, it is 0\%.}
\label{d_vals}
\end{figure}
\begin{figure}[!h]
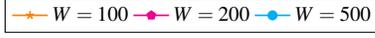

    \centering
    \resizebox{0.6\columnwidth}{!}{\pgfplotslegendfromname{combinedlegend} 
    }
    \caption{Mutual Legend for Figures~\ref{d_vals} and~\ref{benign_analysis_fig}.}
\end{figure}
\begin{figure}[!h]
    \centering
    \resizebox{0.6\columnwidth}{!}{
\begin{tikzpicture}
    \begin{axis}[
        xlabel={\( d (\#hashes) \)},
        ylabel={FP},
        xmin=1.5, xmax=5.5,
    ymin=0.0176, ymax=0.01771,  
        xtick={2, 3, 4, 5},
        scaled ticks=false,
                tick label style={/pgf/number format/fixed,/pgf/number format/precision=5}, 
        ytick={ 0.0176, 0.01765, 0.0177}, 
                y tick label style={rotate=20,anchor=east}, 
        legend style={at={(1.1,1)},anchor=north west}, 
        legend columns=3,
        ymajorgrids=true,
        grid style=dashed,
        width=\columnwidth, 
        height=0.6\columnwidth, 
        legend to name=combinedlegend 
    ]
    
    \addplot[
        color=orange,
        mark=star,
        thick
    ]
    coordinates {
        (2, 0.017683) (3, 0.017631) (4, 0.017613) (5, 0.017610)
    };
    \addlegendentry{\( W = 100 \)}
    
    \addplot[
        color=magenta,
        mark=pentagon*,
        thick
    ]
    coordinates {
        (2, 0.017651) (3, 0.017617) (4, 0.017610) (5, 0.017610)
    };
    \addlegendentry{\( W = 200 \)}
    
    \addplot[
        color=cyan,
        mark=oplus*,
        thick
    ]
    coordinates {
        (2, 0.017610) (3, 0.017610) (4, 0.017610) (5, 0.017610)
    };
    \addlegendentry{\( W = 500 \)}
    
    \end{axis}
\end{tikzpicture}
}
\caption{\textbf{FP as a Function of the $d$ Value - Benign Data}. Different $d$ values for CMS, and their FP rate for $w$ values. It is shown that the largest FP rates are less than 2\%, decreasing to 0\% with 5 hash functions ($d=5$).}
\label{benign_analysis_fig}
\end{figure}
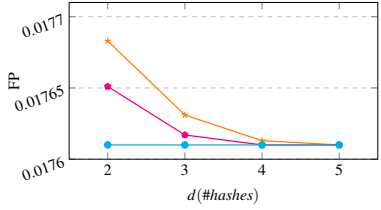

\section{Results}
\label{sec:results} 
\Harel{In this section, we present the results of~\system.}
\subsection{R$\ell$1 Experiments}
\label{r1_exp}


Figure~\ref{d_vals} shows the FP rates of the attack, which exceed 20\% for a window size of $w=100$. For $w=200$, the FP rate is 6\% with two hash functions ($d=2$), decreasing to less than 1\% with five hash functions ($d=5$); for $w=500$, the FP rate is 0\%. Across all experiments, the attack success rate (ASR) was 0.0076\% and thus was not visualized. This low ASR was since after $\tau=5$ packets, all subsequent packets with the same malicious domain within a one-second window were identified as malicious. Note that when a suspicious packet is identified by~\system, a truncated packet is sent to the resolver, prompting it to open a TCP connection to the authoritative server. \Harel{Only valid packets (based on the TXID and port) will be processed.} Thus, although we document an FP rate, there is no actual loss for benign requests wrongly identified as malicious; instead, the resolver obtains the result via TCP and transfers the response to the client.
\subsection{R$\ell$2 \& R$\ell$3 Experiments}
\system~detects the fragmentation and out-of-bailiwick attacks, respectively. As each rule targets a single packet (R$\ell$2 on the first fragment, and R$\ell$3 on the out-of-bailiwick packet), the system found this packet instantly. 

\subsection{Begin Data Experiment} \Harel{We analyzed the FP rate of \system\ on clean, benign data (i.e., without any attacks). Figure~\ref{benign_analysis_fig} illustrates the trends in these FP rates. As observed, for $w=100$, the FP rate on benign data is approximately 2\% across various values of $d$. However, when $d=500$, the FP rate drops to 0\% regardless of the number of hash functions used. As discussed in Section~\ref{r1_exp}, this FP rate is innocuous, as the resolver successfully retrieves the result via TCP and forwards the response to the client.}

In summary, the experiments demonstrated that \system\ provides precise analysis of benign traffic and DNS cache poisoning attacks. Using a configuration of $w=200$ and $d=5$, we achieved an ASR of $0.0076\%$ and an FP rate of $2\%$ on clean data, and $1\%$ on interleaved benign and malicious traffic. To get a 0\% FP, $w=500$ and $d=2$ can be utilized.

\section{\system~as a Proactive CVE Mitigation}
\label{sec:cves}
 In this section, we present 14 CVEs related to targeted DNS software that \system\ prevents from being exploited. Employing \system\ reduces the necessity of fixing resolvers that use vulnerable DNS software mentioned in these CVEs. We compiled the list of CVEs from NIST~\cite{NIST} by searching for "DNS cache poisoning"\cite{nist_dnscachepoisoning_search} and "DNS spoofing"\cite{nist_dnsspoofing_search}. We filtered out CVEs relating to DNS clients, as we did with logic DNS cache poisoning. We also omitted cases where the CVE mentions DNS cache poisoning/DNS spoofing as a tool for another attack without explicitly explaining how this attack is implemented (e.g., attacks on weak TLS/SSL certificates~\cite{nvd_cve_2008_3280}). Additionally, we discarded CVEs related to DNSSEC, as its utilization is out of the scope of this work due to poor distribution and reliance on authoritative servers.
 
 We manually explored the given information and mapped which type of attack can exploit this CVE (based on the types in Section~\ref{attack_types}). \Harel{Through this mapping, we demonstrate that~\system~mitigates all reported vulnerabilities associated with the following CVEs, as it has previously addressed the corresponding types of attacks.} We also described the vendor affected by each CVE.
 Then, we documented the rule of~\system~that captures this CVE.
 Following is a brief description of each CVE:
 \begin{enumerate} \item \textbf{CVE-2023-30464~\cite{nvd_cve_2023_30464}} is a recent CVE affecting CoreDNS, a Golang-based DNS server library. The exploitation involves sending multiple attacks with varying TXID and source port values, leveraging the recent Tudoor attack~\cite{li2023tudoor}. This exploit is thus classified as an $S$ attack, similar to Tudoor.
 \item \textbf{CVE-2023-28457~\cite{nvd_cve_2023_28457}} was also discovered through the Tudoor attack. In this case, vulnerabilities were found in Technitium DNS and Microsoft DNS, susceptible to similar exploitation as CVE-2023-30464. This vulnerability is similarly classified as an $S$ attack.
 
 \item \textbf{CVE-2023-28457~\cite{nvd_cve_2021_43105}} reveals a vulnerability in Technitium DNS that allows injection of NS records, even at the TLD level, due to improper validation of the bailiwick rule. This CVE is exploited through a $B_{OOB}$ attack, as it involves a packet that violates the bailiwick rule.
 
 \item \textbf{CVE-2021-3448~\cite{nvd_cve_2021_3448}} discloses a vulnerability in dnsmasq, enabling an attacker on the network to exploit a fixed port used in query forwarding, facilitating DNS cache poisoning by guessing random TXIDs. This CVE aligns with an $S$ attack due to the guessing of TXIDs.
 
 \item \textbf{CVE-2020-25684~\cite{nvd_cve_2020_25684,DNSpooq}} identifies an anomaly in dnsmasq, where multiple unanswered queries are forwarded to an upstream server using only 64 random source ports. With multiple TXIDs sharing a port, the effective entropy is reduced to 26 bits, enabling attackers to guess the TXID-port combination through brute force, as seen in $S$ attacks. Cisco VPN routers and OpenWRT firmware were among the affected platforms.
 
 \item \textbf{CVE-2017-12132~\cite{nvd_cve_2017_12132}} reveals a vulnerability in the DNS resolver within the glibc library. Exploitation uses large UDP responses, leading to fragmentation, and is categorized as an $S_{Frag}$ attack.
 
 \item \textbf{CVE-2008-3217~\cite{nvd_cve_2008_3217}} describes a flaw in source port randomization in PowerDNS, enabling TXID brute-force attacks. This CVE is thus mapped to an $S$ attack.
 
 \item \textbf{CVE-2008-1447~\cite{nvd_cve_2008_1447}} highlights vulnerabilities in the random generation of source ports and TXIDs in BIND and Microsoft DNS, classifying it as an $S$ attack.
 
 \item \textbf{CVE-2008-1454~\cite{nvd_cve_2008_1454}} documents a vulnerability in Microsoft DNS that allows the resolver to retrieve records outside of the delegated authoritative domain. This vulnerability is exploited via an outside-of-bailiwick rule attack, categorized as $B_{OOB}$.
 
 \item \textbf{CVE-2008-1146~\cite{nvd_cve_2008_1146}} describes Klein's attack~\cite{klein2007bind} on OpenBSD's BIND, involving guessing 10 TXIDs for a single query, thus making it an $S$ attack.
 
 \item \textbf{CVE-2007-2926~\cite{nvd_cve_2007_2926}} identifies a weak random number generator for TXIDs, which is susceptible to brute-force $S$ attacks.
 
 \item \textbf{CVE-2002-2211~\cite{nvd_cve_2002_2211}} documents a vulnerability in BIND where attackers can send multiple queries for a single resource record, along with spoofed responses. The attack involves sending multiple responses per query, fitting the $S$ attack classification.
 
 \item \textbf{CVE-2002-2212~\cite{nvd_cve_2002_2212}} reports the same vulnerability as CVE-2002-2211, but for Fujitsu UXP/V.
 
 \item \textbf{CVE-2002-2213~\cite{nvd_cve_2002_2213}} reports the same vulnerability as CVE-2002-2211, affecting Infoblox DNS.
\end{enumerate}
 

 Table~\ref{tab:cve_tab} illustrates the CVEs. Similar to the case of the logic DNS cache poisoning attack, rule R$\ell$1 matches the majority of CVEs (78\%), followed by R$\ell$3 (14\%) and R$\ell$2 (7\%).

\begin{table}[!h]
    \centering
    \begin{tabular}{ccp{2cm}||c}
        \toprule
        \textbf{CVE} & \textbf{Type} &
        \textbf{Vendor} &
        \textbf{\system} \\
        \midrule
        CVE-2023-30464~\cite{nvd_cve_2023_30464} &
        $S$&CoreDNS &R$\ell$1 \\ \hline
        CVE-2023-28457~\cite{nvd_cve_2023_28457} & $S$&Microsoft DNS, Technitium&R$\ell$1\\ \hline
        CVE-2021-43105~\cite{nvd_cve_2021_43105} & $B_{OOB}$&Technitium&R$\ell$3 \\ \hline
        CVE-2021-3448~\cite{nvd_cve_2021_3448} & $S$&dnsmasq&R$\ell$1 \\ \hline
        CVE-2020-25684~\cite{nvd_cve_2020_25684,DNSpooq} & $S$&dnsmasq,Cisco, OpenWRT &R$\ell$1 \\ \hline
        CVE-2017-12132~\cite{nvd_cve_2017_12132} & $S_{Frag}$&-&R$\ell$2 \\\hline
        CVE-2008-3217~\cite{nvd_cve_2008_3217} & $S$&PowerDNS&R$\ell$1 \\ \hline
        CVE-2008-1447~\cite{nvd_cve_2008_1447} & $S$&BIND, Microsoft DNS&R$\ell$1 \\ \hline
        CVE-2008-1454~\cite{nvd_cve_2008_1454} & $B_{OOB}$&Microsoft DNS&R$\ell$3 \\ \hline
        CVE-2008-1146~\cite{nvd_cve_2008_1146} & $S$&OpenBSD's BIND&R$\ell$1 \\ \hline
        CVE-2007-2926~\cite{nvd_cve_2007_2926} & $S$&ISC BIND&R$\ell$1 \\ \hline
        CVE-2002-2211~\cite{nvd_cve_2002_2211} & $S$&BIND&R$\ell$1 \\ \hline
        CVE-2002-2212~\cite{nvd_cve_2002_2212} & $S$&Fujitsu UXP/V&R$\ell$1 \\ \hline
        CVE-2002-2213~\cite{nvd_cve_2002_2213} & $S$&Infoblox DNS&R$\ell$1 \\ 
        \bottomrule
    \end{tabular}
    \caption{\textbf{CVEs fixed by~\system.} For each CVE, we describe the attack type used to exploit it based on its description, document the vendors affected by the vulnerability, and list the corresponding rule of \system~ that matches the CVE.}
    \label{tab:cve_tab}
\end{table}

\section{Comparison to Other Tools}
\label{sec:tools}
\Harel{In this section we compare~\system~to other IDS/IPS tools, Suricata and Snort.}
\textbf{Suricata}~\cite{suricata-website} is an open-source IDS/IPS and NSM engine that uses multi-threading and signature-based detection to monitor network activity. We compare~\system~to Suricata's DNS rules~\cite{suricata-1,suricata-2} from the OpenWRT ruleset, as the free version available on the Suricata website~\cite{suricata-website} currently lacks these rules. 
All following Suricata rules are found to be related to DNS cache poisoning:

    \begin{enumerate}
        \item \textbf{SID:2008446} At least 100 DNS responses in 10 seconds. The responses are aggregated by only source IP. The rule also makes sure that there is at least one record in the authority part.
        \item \textbf{SID:2008447,2008475} Kaminsky attack identification. These rules match sequences of A/NS responses of at least 50 responses in 2 seconds per source IP.
        \item \textbf{SID:2013894} A tailored rule of attack on Google Brazil's domain (google.com.br). This rule looks for a set of 100 responses in 10 seconds from the same source IP. Each response should contain one record at least in the answer section and in the authority section.
        \item \textbf{SID:2100253,2100254} Low TTL response. These rules look for responses with no authority section. The first rule specifically looks for PTR response.
    \end{enumerate}

\Harel{The last two Suricata rules were not compared to our system. Rules 2013894, 2100253, and 2100254 did not provide sufficient packet details to enable a direct comparison with our system. It’s important to note that the aggregation of Suricata's rules is based on IP addresses, while R$\ell$1 performs aggregation based on subdomains, with a timing threshold of 1 second (compared to Suricata’s 2 seconds).}

\textbf{Snort}\cite{koziol2003intrusion,caswell2007snort} is an open-source IDS/IPS, now maintained by Cisco, that uses a signature-based approach to detect and respond to network threats. To compare with~\system, we identified Snort rules targeting DNS cache poisoning attacks, which can be found by searching "DNS cache" on the Snort website and examining the "snort3-protocol-dns.rules" file from the latest subscribed ruleset. 
Some of the rules were not detailed, and others were disabled by Snort, so they cannot be compared to~\system. Therefore, we describe four still-running rules with explicit details of DNS cache poisoning attacks:

        

\begin{enumerate}
    \item \textbf{SID 2008446:} 
    This rule is triggered when the DNS response packet contains more than zero resource records (RRs) in both the answer section and the additional section, with over 100 such packets observed from the same source IP within 10 seconds.

    \item \textbf{SID 2008447:} 
 This rule indicates a DNS response indicating 3 RRs. The rule is triggered if there are more than 50 such packets from the same source within 2 seconds. 
    \item \textbf{SID 2008475:} 
     This rule indicates 3 A RRs. It is triggered if more than 50  packets are observed from the same source within 2 seconds. The rule aims to detect potential A RR cache poisoning attempts.

    \item \textbf{SID 2008446:} The same as Suricata's SID:2013894 rule. 
    
\end{enumerate}

The last rule was not compared, similar to the equivalent Suricata rule. Also, the aggregation of Snort's is similar to Suricata's, which is less effective than our system. 

\Harel{In summary, we found that both Suricata and Snort provide rules that somewhat resemble our R$\ell$1 rule; however, they aggregate packets based on IP addresses, while our approach aggregates based on subdomain names. By focusing on specific subdomains, our method captures attacks more efficiently. Furthermore, their rules trigger after a minimum of 2 seconds and 50 packets, whereas our tool operates with a 1-second threshold and just 5 packets, enabling faster detection. Additionally, their rules do not account for fragmentation attacks, which our tool is designed to capture.}

\section{Discussion}
\label{sec:diss}
Our new system, \system, assumes that an attacker follows one of three methods—extensive traffic per one query searching for port/TXID, fragmentation, or "out of bailiwick". In the past, a study by Klein et al.\cite{klein2021cross} presented an attack that assumed the attacker knows the port or TXID from an external source (beyond the DNS protocol itself). Based on this past work, an attacker can adopt a new method: first, the attacker identifies the source port the resolver uses through unspecified techniques. Then, for multiple distinct queries, the attacker sends a single fake response for each, in a packet that adds an authority record and an additional record inside the bailiwick. This way, the attacker evades our rules. By targeting hours with low traffic to the DNS resolver, the attacker can achieve a high Attack Success Rate (ASR). The attacker can also use other templates\cite{klein2017internet} as well as traditional NS/additional records.

Indeed, this attack does not trigger our rules. However, we conducted an assessment of Tranco’s top 1000 domains~\cite{pochat2018tranco} using simple DNS queries and analyzed the distribution of response types, categorizing them as follows: 46\% of the responses included only answer records, 26\% included both answer and authority records, 22\% had no answer records (but had additional or authority records), and 6\% featured answer, authority, and additional records. This preliminary assessment indicates that DNS responses exhibit a typical statistical distribution that can be monitored, and while an attacker could potentially exploit this statistical behavior, the method would require strict adherence to these patterns. A more in-depth analysis would be necessary to establish an additional statistical rule for inclusion in~\system, and this is left for future work.

\section{Conclusion}
\label{sec:conc}
In this study, we introduced \system, a system designed to protect resolvers against various DNS cache poisoning attacks and mitigate exploits stemming from CVEs. We demonstrated that \system~is more effective for this task compared to Suricata and Snort. Its retrospective capability captures known DNS cache poisoning attacks and provides a foundation for mitigating future threats, similar to Akamai's recent announcement~\cite{akamai2024keytrap}. For instance, an attack that leverages a combination of fake second fragments and packets scanning source ports (2013B + 2021A) to maximize impact can be effectively detected through the combination of R$\ell$1 and R$\ell$2. Our simulation of real-world data further validated the system's performance by assessing the rate of false positives generated, thus proving \system’s applicability in practical scenarios.  


We acknowledge several limitations of our work. First, while our analysis focuses on DNS cache poisoning attacks, it does not immediately scale to other types of DNS attacks, such as DDoS. Second, our system cannot identify attacks that involve BGP hijacking or other network protocols combined with DNS cache poisoning. Additionally, attacks that bypass the need to guess both the port and TXID and fall below the detection threshold will evade our system. These limitations present opportunities for future research and potential extensions of \system.
\section{Ethics considerations}
\label{sec:eth}
We evaluated~\system~for false positives (FP) on real data obtained from a published dataset on the Internet\cite{singh2019}. We carefully considered the ethical implications of our analysis and ensured that user privacy was protected. To prevent any harm from using user data in DNS queries, we only retained the responses received by the resolver from DNS authoritative servers, discarding the rest of the data. As a result, no IP addresses or personally identifiable information (PII) from users were used at any stage of the research. \Harel{All the experiments of~\system~were conducted in a controlled laboratory
environment, with dedicated server.} 

\section{Open Science}
\label{sec:open_sc}
\Harel{We fully comply with the USENIX Security 2025 Open Science Policy by openly sharing the research artifacts associated with this work. A comprehensive Git repository containing all code, PCAP files, and results is available as an anonymous Github repository\footnote{https://anonymous.4open.science/r/Poisoning-Prevention-System-EE50}. This repository ensures the reproducibility and replicability of our findings. The full repository will be made permanently available after the paper's acceptance, with no licensing restrictions preventing its dissemination. All artifacts will also be submitted to the Artifact Evaluation Committee for review prior to the final submission deadline.}

\section{Acknoledgements}
We would like to thank Prof. Amit Klein
(The Hebrew University) and Prof. Haya
Shulman (Goethe-Universität Frankfurt) for fruitful discussions. Additionally, we are grateful to Prof. Micah Sherr (Georgetown
University) for providing a supportive environment and the
necessary equipment for this research.  This work was supported in
part by an ISF Grant number 1527/23, and a grant from the Blavatnik Interdisciplinary Cyber Research Center (ICRC), Tel Aviv
University, and the Blavatnik family fund.

\bibliographystyle{ieeetr}
\bibliography{biblio}
\appendix\section{Appendix - Full Analysis - R$\ell$1}
\label{appendix_r1}
\subsubsection{Memory Usage Calculation}

\begin{itemize}
    \item 
\textbf{Count-Min Sketch}  Given the large difference between the ranges, we can use low values for \(d\) and \(w\) to minimize memory usage while still achieving effective differentiation. By setting \(d = 5\) (number of hash functions) and \(w = 200\) (number of counters per row), we get:

\[
w \times d \times \text{size of counter} = 200 \times 5 \times 4 = 4000 \text{ bits}
\]
    
    \item \textbf{Fixed-size dwsHH}: The memory usage is calculated as \( O(k\ell \log m) \). Assuming \( k = 100 \) and \( \ell = 32 \), and \( m \approx 0.1N \):
    \[
    \text{For 10 domains: } 100 \times 32 \times \log(1) = 3.2K \text{ bits}
    \]
    \[
    \text{For 100 domains: } 100 \times 32 \times \log(10) = 3.2K \text{ bits}
    \]
    \[
    \text{For 1K domains: } 100 \times 32 \times \log(100) = 6.4K \text{ bits}
    \]
    \[
    \text{For 10K domains: } 100 \times 32 \times \log(1000) = 9.6K \text{ bits}
    \]
    \item \textbf{Fixed-threshold WS}: The memory usage is calculated as \( O(\tau \sum_y w_y \cdot \ell \log m) \). With \(\tau = 0.01\):
    \[
    \text{For 10 domains: } 0.01 \times 10 \times 32 \times \log(1) = 3.2 \text{ bytes}
    \]
    \[
    \text{For 100 domains: } 0.01 \times 100 \times 32 \times \log(10) = 32 \text{ bytes}
    \]
    \[
    \text{For 1K domains: } 0.01 \times 1K \times 32 \times \log(100) = 640 \text{ bytes}
    \]
    \[
    \text{For 10K domains: } 0.01 \times 10K \times 32 \times \log(10) = 9.6K 
    \text{ bytes}
    \]
\end{itemize}
The comparison of memory usage is presented in Fig.~\ref{fig:memory}.


\subsubsection{Error Rates Calculation}

\begin{itemize}
    \item \textbf{Count-Min Sketch (CMS)}: 
    \[
\epsilon = \frac{e}{w} \approx \frac{2.718}{200} \approx 0.01
\]
    \item \textbf{Fixed-size dwsHH}: The error formula is given by:

\[
\text{Error} = \frac{1}{k} \sum_y w_y + \frac{w_y}{\sqrt{2\ell}}
\]

Assuming \( w_y = 1 \) for all \( y \) and \( \ell = k \), we calculate the error for different cache sizes \( k \):\\
For 10 domains:
$\frac{1}{10} \sum_y 1 + \frac{1}{\sqrt{2 \times 10}} = \frac{10}{10} + \frac{1}{\sqrt{20}} =  1 + \frac{1}{4.47} \approx 1 + 0.223 = 1.223$\\
For 100 domain:
$\frac{1}{100} \sum_y 1 + \frac{1}{\sqrt{2 \times 100}} = \frac{100}{100} + \frac{1}{\sqrt{200}} = 1 + \frac{1}{14.14} \approx 1 + 0.071 = 1.071
$\\
For 1000 domains:
$\frac{1}{1000} \sum_y 1 + \frac{1}{\sqrt{2 \times 1000}} = \frac{1000}{1000} + \frac{1}{\sqrt{2000}} = 1 + \frac{1}{44.72} \approx 1 + 0.022 = 1.022$\\
For 10000 domains:
$\frac{1}{10000} \sum_y 1 + \frac{1}{\sqrt{2 \times 10000}} = \frac{10000}{10000} + \frac{1}{\sqrt{20000}} = 1 + \frac{1}{141.42} \approx 1 + 0.007 = 1.007
$

    \item \textbf{Fixed-threshold WS}: The error bound is calculated as \(\tau^{-1} + \frac{w_y}{\sqrt{2\ell}}=0.01^{-1} + \frac{1}{\sqrt{64}} =  100.125\):
    
\end{itemize}
The comparison of error rates is presented in Fig.~\ref{fig:error}.

\subsubsection{Inference Time Calculation}

\begin{itemize}
    \item \textbf{Count-Min Sketch (CMS)}: The query time is \( O(d) \), with \( d = 5 \), involving 5 hash lookups.
    \item \textbf{Fixed-size dwsHH}: The query time is \( O(\log n) \) for keys not in cache, and 2 accesses for keys in the cache.
    \item \textbf{Fixed-threshold WS}: The query time is \( O(\tau N) \):
    \[
    \text{For 10 domains: } 0.01 \times 10 = 0.1
    \]
    \[
    \text{For 100 domains: } 0.01 \times 100 = 1
    \]
    \[
    \text{For 1000 domains: } 0.01 \times 1000 = 10
    \]
    \[
    \text{For 10K domains: } 0.01 \times 10000 = 100
    \]
\end{itemize}

\section{Appendix - Full Algorithm}
\label{appendix_algo}

\begin{algorithm}
\caption{\system}
\label{alg:main}
\begin{algorithmic}[1]
\Require Set of packets $P = \{p_1, p_2, \ldots, p_n\}$, Count-Min Sketch (CMS), Threshold $\tau$, Check interval $N$, Counter $count$, Window size $W$, initial time $t$, domain map $domainMap$, global $action$
\Ensure Global constants: $CMS, \tau, N$
\For{each packet $p_i \in P$} \Comment{Iterate through all packets}
    \If{$R\ell1(p_i,domainMap)$ \textbf{or} $R\ell2(p_i)$ \textbf{or} $R\ell3(p_i)$}
    \If{$p_i!=NULL$}
    \State $action \gets "truncate"$
        \State $p_i.\text{TC} \gets 1$ \Comment{Set Truncated bit}
    \State $p_i.\text{Ans} \gets \text{blank}$ \Comment{Clear answer record}
    \State $p_i.\text{Auth} \gets \text{blank}$ \Comment{Clear auth record}
    \State $p_i.\text{Add} \gets \text{blank}$ \Comment{Clear add record}
    \EndIf
    \Else
    \State $action \gets "forward"$ \Comment{No rule fired}
    \EndIf

    \State Send $p_i$ to $p_i.\text{dest}$ \Comment{Send packet to its destination}
\EndFor

\end{algorithmic}
\end{algorithm}

\begin{algorithm}
\caption{\textsc{R$\ell$1} Function - Probabilistic Count of Domains}
\label{alg:Ru1}
\begin{algorithmic}[1]
\Function{R$\ell$1}{$p,domainMap$}
    \Comment{Initialize result as False}
    \State CMS.Add($p.\text{domain}$) \Comment{Update CMS with  $p$}
    
    \If{$count \mod N == 0$} \Comment{Check frequency}
        \If{CMS.Freq($p$) $> \tau$}
        \If{$domainMap[p.domain]$ == true}
        \State \Return true
        \Else
            \State $domainMap[p.domain] \gets true$ \Comment{Signal the domain is from now suspicious}
            \State \Return true \Comment{Now, truncate the packet}
        \EndIf
    \EndIf
    
    \State $count \gets count + 1$ \Comment{Increment packet counter}

    \If{$p.{Time}-t \mod W > 0$} \Comment{window t/o}
        \State $count \gets 0$
        \Comment{init packet counter}
        \State $CMS \gets New(CMS)$
        \Comment{init the CMS table}
        \State $domainMap \gets domainMap$
        \Comment{init map}
        \State $t \gets p.{Time}$
        \Comment{Init the time variable}
    \EndIf
\EndIf
    \State \Return false
\EndFunction
\end{algorithmic}
\end{algorithm}

\begin{algorithm}
\caption{\textsc{R$\ell$2} Function - Check for Fragmentation}
\label{alg:Ru2}
\begin{algorithmic}[1]
\Function{R$\ell$2}{$p$}
\Comment{Initialize result as False}    
    \If{$p$.offset $== 0$ $\&$ $p$.MF == \textbf{true}} 
    \Comment{1st frag}
        \State \Return true
    \Else
    {$p$.offset $> 0$} 
    \Comment{2nd frags}
        \State $p \gets NULL$
    \EndIf
    \State \Return false
\EndFunction
\end{algorithmic}
\end{algorithm}
\begin{algorithm}
\caption{\textsc{R$\ell$3} Function - Bailiwick Rule Compliance}
\label{alg:Ru3}
\begin{algorithmic}[1]
\Function{R$\ell$3}{$p$}
    \State $is\_valid \gets \textbf{true}$
    
    \ForAll{$record \in p.answers$}
        \If{\textbf{not} \Call{IsWithinBailiwick}{$record.name, p.queryname$}}
            \State $is\_valid \gets \textbf{false}$
            \State \textbf{break}
        \EndIf
    \EndFor
    
    \If{$is\_valid$}
        \ForAll{$record \in p.authoritative$}
            \If{\textbf{not} \Call{IsWithinBailiwick}{$p.queryname,record.name$}}
                \State $is\_valid \gets \textbf{false}$
                \State \textbf{break}
            \EndIf
        \EndFor
    \EndIf
    
    \If{$is\_valid$}
        \ForAll{$record \in p.additional$}
            \If{\textbf{not} \Call{IsWithinBailiwick}{$record.name, p.queryname$}}
                \State $is\_valid \gets \textbf{false}$
                \State \textbf{break}
            \EndIf
        \EndFor
    \EndIf
    \If{$\neg is\_valid$} \Return true \Else~\Return false
    \EndIf
    
\EndFunction

\Function{IsWithinBailiwick}{$domain, origin$}
    \State $is\_within \gets \textbf{false}$
    \If{\Call{EndsWith}{$domain, origin$}}
        \State $is\_within \gets \textbf{true}$
    \EndIf\\
    \Return $is\_within$
\EndFunction

\end{algorithmic}
\end{algorithm}

\end{document}